# Kinetics of excitations on the Fermi arcs in underdoped cuprates at low temperature

Lev P. Gor'kov
*NHMFL, Florida State University, 1800 East Paul Dirac Drive, Tallahassee Florida 32310, USA
and L.D. Landau Institute for Theoretical Physics of the RAS, Chernogolovka 142432, Russia*

The Fermi-liquid-like (FL) resistivity recently observed in clean Hg1201 below the pseudogap temperature was related to carriers at the nodal points on the Fermi surface (FS) [4]. We show that this necessitates important implications for the electronic spectrum of UD cuprates in whole. Photoemission experiments picture the spectrum as of "metallic arcs" separated from each other by regions with large energy gaps. We solved the kinetic equation in such model rigorously. The Fermi arcs' carriers contribute to the FL resistivity, if scattering between the opposite nodal points admits the Umklapp processes. The Hall coefficient defines the effective number of carriers on arcs and has the positive sign. For clean materials the expression is applicable only at weak magnetic fields. We discuss the Fermi arcs' concept further in light of recent experimental findings and argue that the idea of reconstructed FS in UD cuprates is not consistent with the FL-like resistivity.

High temperature superconductivity (HTSC) in cuprates challenges the condensed matter physics community for already more than twenty five years. Besides that the pairing mechanisms are unknown, even in the paramagnetic phase cuprates demonstrate strange non-Fermi liquid energy spectrum. Thus, for UD cuprates photoemission experiments find the coherent excitations only within narrow "arcs" at the nodal points of the "bare" FS (for review, see [1]). While general consent is that responsible for non-Fermi liquid behavior are strong electron-electron correlations in vicinity of the Mott insulating antiferromagnetic phase, a unifying theory is still missing.

Recent developments concerned discovery of the quantum oscillations (QOs) in the vortex state that revealed small Fermi liquid (FL) pockets in UD *ortho II* YBCO [2]. In literature interpretation of these findings varies from a scenario of the *two* pockets formed on the "bare" FS reconstructed in a hypothetical spin- or density-wave phase transition at higher temperatures [2] and the idea of *single* pocket inherent in details of the band structure [3].

Most recently one more feature was reported on for Hg1201 [4]. Besides $T_c$, the temperature of superconducting transition and $T^*$, the pseudogap temperature, there exists a new scale, $T^{**}$ on the phase diagram of Hg1201 below which resistivity manifests the distinct $T^2$-dependence typical for FL [4]. An interval of the order of $T^{**}$ itself separates $T^{**}$ and $T^*$ from each other. Above $T^*$ resistivity is proportional to temperature, as usual in the pseudogap regime in cuprates.

It was emphasized in [4] that the new feature is not unique for Hg1201 and is seen in the resistivity data available for a few other UD cuprates (YBCO and LSCO). Consequently, the question arises whether the emergent $T^2$-resistivity regime [4] in Hg1201 has something in common with small pockets first deduced in [2] from low frequency quantum oscillations in *ortho II YBCO* in high magnetic fields.

Photoemission (ARPES) and the magneto-transport studies are the two independent experimental techniques. In UD cuprates ARPES sees *single arcs* (the Fermi arcs) at each of the four nodal points in the Brillouin zone (BZ), while reconstruction of the FS via a spin- or charge-density wave transition hypothesized first in [2] is expected to result in *two small pockets*, at $(0, \pm\pi), (\pm\pi, 0)$ and $(\pm\pi/2, \pm\pi/2)$ points (see recent review [5]). In what follows, we demonstrate that the $T^2$-FL resistivity [4] together with the results [3] give firm support to the concept of the Fermi *arcs*.

Interpretation in [2, 6] of QOs as due to two pockets on the reconstructed FS suffers from the implicit reliance that superconductivity is suppressed in such surprisingly low fields as $H \approx 28\text{-}30T$. That is not necessary for possible observation of QOs. Indeed, QOs in the vortex state are common to many multi-band superconductors where SC is driven by strong correlations on one or more "main" FSs, while the gaps induced on the rest are small [7].

In [3] the electronic specific heat in UD *ortho II YBCO* was studied as in the presence of the magnetic field, as in the field absence, and down to low temperatures (>4K). Notable, since the analysis of QOs together with the oscillation's frequency that gives size of the pocket provides the carriers' mass as well, the linear in temperature electronic contribution to the specific heat *at zero fields* could be calculated directly. The Sommerfeld coefficient obtained in this way coincided with its experimental value [3] at low $T$ in the field absence. From this unequivocally follows existence of only *one* pocket.

Self-evidently, the Cooper instability must be directly related to pairing of excitations on the opposite arcs. Then, should electrons on the "arcs", in point of fact, form closed pockets, the onset of superconductivity would build on the latter the *d*-wave superconducting gap below $T_c$, contrary to the linear in $T$ [3] electronic specific heat at low temperature in *zero fields*. In turn, being outside the nodal points the pocket [3] is too small in size to admit the Umklapp processes and cannot contribute to finite conductivity. Thereby, carriers at the Fermi arcs and at the small pocket belong to two different sub-systems.

Analysis of the Drude formula for resistivity $\rho(T) = [m*/ne^2 \tau(T)]$ reveals where the peculiarity of the new regime lies [4]. In layered cuprates the FL $T^2$-contribution can conveniently be presented as:

$$(s/c)(e^2/\pi\hbar)\Delta\rho(T^{**}) = const \times \left(T^{**}/\varepsilon_F\right)^2. \qquad (1)$$

($c$ is the lattice constant perpendicular to the $CuO_2$-plane; $s$-the number of conducting layers per the unit cell).

In theory, the $T^2$-dependence signifies the FL regime as long as $T \ll \varepsilon_F$ in the whole temperature interval, i.e., the Fermi energy *must be large*. Substitution of $T^{**}$ and use of the resistivity data e. g., for YBCO (see Fig. 3[4]) gives: $\varepsilon_F \approx (const)^{1/2} 290K$ at $T^{**} = 190K$ (dimensionless $const. \sim 1$). The fact that the quadratic dependence describes resistivity of YBCO up to $T^{**} \approx (2/3)\varepsilon_F$ puts the straightforward FL interpretation Eq. (1) in question.

After searching through various mechanisms, from the charge-wave order, superconducting fluctuations to the FL regime via the hypothetic pocket, such as was deduced from QO in UD YBCO, the authors [4] associated the $T^2$ regime with carriers at the nodal points on the Fermi surface.

Although qualitative arguments [4] SI (9) is somewhat vague, we adopted this point of view. Below we suggested the model of the Fermi arcs and solved it rigorously. The $T^2$- resistivity regime at low enough temperatures is obtained after taking explicitly into account the Umklapp scattering between the opposite nodal points. Forestalling results of the analysis below, the physics behind, loosely speaking, is that it is the total number of *carriers on the Fermi arcs* that must be taken as the number $n$ of charge carriers in the Drude formula. After that, the Fermi energy in Eq. (1) turns out large: $\varepsilon_F^* = p_F^2/2m^*$, of the order of the chemical potential ($m^*$ is the renormalized band mass).

Before to proceed further, note in passing that discussions whether ARPES technique distinguishes between "arcs" and "pockets" continued until recently (e.g., [8]). We adopt the view that arcs are quite literary "single-sided arcs" without "hidden contours" on their backside.

In early experiments [9] the arc's length was found proportional to temperature at all doping so that at extrapolation to $T = 0$ the arc would collapse onto single point. Results [9] were revised in [10-12] and it is now well established that the length of arcs' in the pseudogap phase is finite down to the lowest temperatures, although the arc shrinks with decrease of the dopants concentration. ARPES reveal the large energy gaps outside the arcs' end points in the normal phase *directly above* $T_c$ [11-13].

One must stipulate that most ARPES data cited above were obtained on samples of the two-layer $Bi_2Sr_2CaCu_2O_8$ (Bi2212), the single-layer Bi2201 and LSCO (see short review [13]). Materials' surface problems continue to plague ARPES experiments in Hg1201 and especially in YBCO [14]. Yet, the general consensus is that except possibly minor details the basic physics behind these ARPES results is of the general character.

With that in mind, we suggest that in clean UD cuprates at low temperature, but still in the normal phase, the energy spectrum of excitations inside the Fermi arc is the ordinary $\varepsilon(\vec{p}) = v_F(p - p_F)$. (Actually, there is a structure in the energy spectrum seen, in particular, as kinks for the Fermi velocity, $v_F$ near the chemical potential; $v_F$ changes only by an insignificant numerical factor [15]). The angular dependence $(t_\perp / 2)|\cos(k_x a) - \cos(k_y a)|$ of large energy gap (pseudogap) that grows from the arc's ending points resembles the profile of the *d*-wave superconducting gap. In UD cuprates $t_\perp$ is considerable larger than $\Delta_0$, the amplitude of the superconducting *d*-wave gap [11-13].

ARPES intensity is proportional to the single-particle spectral function $A(\omega, \vec{k})$ (see e.g., [1]). The latter is related in the following way to the imaginary part of the retarded Green function, $G(\vec{k}; \omega) = [\omega - \varepsilon(\vec{k}) + \mu - \Sigma(\vec{k}; \omega)]^{-1}$:

$$A(\omega, \vec{k}) = -\frac{1}{\pi} \frac{\mathrm{Im}\Sigma(\vec{k}; \omega)}{[\omega - \varepsilon(\vec{k}) + \mu - \mathrm{Re}\Sigma(\vec{k}; \omega)]^2 + [\mathrm{Im}\Sigma(\vec{k}; \omega)]^2}. \qquad (2)$$

In the experimental procedure (e.g., [9]) positions of excitations on the ungapped parts of the "bare" FS, in accordance to Eq. (2), are expected to coincide with maxima of $A(\omega, \vec{k})$ at $\omega = 0$. Inside each arc the Green function can be written in the customary form as first obtained for the electrons interacting with phonon in metals [16] and for the "Kondo lattice" problem (in the mean field approximation) [17]:

$$G^{R(A)}(\vec{p}; \omega) = \frac{Z}{\omega - Z[\varepsilon(\vec{p}) - \mu] \pm i/\tau(\omega)} \cong \frac{Z}{\omega - Z\xi \pm i/\tau(\omega)}, \qquad (3)$$

(Here $\xi = v_F(p - p_F)$ is the perpendicular to FS component of the "bare" energy dispersion, $\tau(\omega)$ - the relaxation time). In (3) the sign (+) or (-) in front of the imaginary part stands for the retarded (*R*) and the advanced (*A*) Green functions, correspondingly. The residue is: $Z = [1 - \partial\Sigma(\omega)/\partial\omega]^{-1}$; $m^* = Z^{-1}m$ - the effective mass [16].

Temperature dependence of the imaginary part $\mathrm{Im}\Sigma(\vec{p}; \omega)$ can be obtained *rigorously* by the analytical continuation of the "skeleton" self-energy diagram with three internal electronic lines [18].

While this is all that one needs to know for the spectral function $A(\omega, \vec{k})$ Eq. (2), is not enough for finding conductivity. Indeed, contributions into $\text{Im}\Sigma(\vec{p};\omega)$ come about from all scattering channels, whereas the conductivity in metals is finite exclusively due to the Umklapp processes in which electrons transfer the momentum to the lattice. Although a *large* enough FS admits the Umklapp scattering, the imaginary part, $\text{Im}\Sigma(\vec{k};\omega)$ determines contributions relevant to conductivity only *by the order of magnitude*. Therefore, to calculate the conductivity in the model of small arcs directly we resort to the kinetic equation. In hindsight the rigorous Green functions formalism allows to express the final results via renormalized masses and strength of interactions.

The kinetic equation reads:

$$e(\vec{E} + \frac{1}{c}[\vec{v} \times \vec{H}])\vec{\nabla}_{\vec{p}} n(\vec{p}) = (\frac{dn}{dt})_{coll}. \quad (4)$$

where $n(\vec{p})$ is the distribution function for carriers with charge $e$, $\vec{E}$ and $\vec{H}$ are the electric and the magnetic fields, correspondingly. Consensus is that strong electron-electron correlations determine fundamentals of the cuprates' physics. Correspondingly, at least, as far as phonons with large momenta are "frozen out" at lower temperatures the kinetics will be due to strong Coulomb (*e-e*) interactions as well.

The textbook integral $(dn/dt)_{coll}$ balances "gains" and "losses" at the (*e-e*) collisions:

$$(\frac{dn_{\vec{p}}}{dt})_{coll} = 2\iint |V(\vec{p}_1 \vec{p}_3; \vec{p}\vec{p}_2)|^2 \frac{d^2\vec{p}_1 d^2\vec{p}_2}{(2\pi)^4} \delta(\varepsilon_1 + \varepsilon_3 - \varepsilon - \varepsilon_2)\{(gains) - (losses)\}, \quad (5)$$

where $\varepsilon_i \equiv \varepsilon_i(\vec{p}_i)$; "*gains*" $= n(\vec{p}_1)n(\vec{p}_3)[1 - n(\vec{p})][1 - n(\vec{p}_2)]$; expression for "*losses*" follows from obvious permutations.

Assuming that Fermi-arcs in UD cuprates are located not far from four points $(\pm\pi/2, \pm\pi/2)$ in the BZ, consider the system in a weak electric field, $\vec{E}$ along the diagonal $(-\pi,-\pi)-(\pi,\pi)$. Then the contribution into the current from the pair of nodal points on the perpendicular diagonal can be neglected. Substitution $\varepsilon(\vec{p}) \to \varepsilon(\vec{p}) - \vec{p}\cdot\vec{u}$ in $n_0(\vec{p}) = (e^{\varepsilon(\vec{p})/T} + 1)^{-1}$ ($\vec{u}$ -the drift velocity) gives the Fermi distribution in the presence of a non-zero current. Expanding over small $\vec{u}$, $n(\vec{p}) \approx n_0(\vec{p}) + n_1(\vec{p})$ for $n_1(\vec{p})$ one has:

$$n_1(\vec{p}) = (\vec{p}\cdot\vec{u})(\frac{1}{4T})ch^{-2}(\frac{\varepsilon_p}{2T}). \quad (6)$$

Conservation law $p^{\|} + p_2^{\|} = p_1^{\|} + p_3^{\|} - K^{\|}$ for the momentum components along the diagonal invokes the Umklapp vector, $\vec{K} = (2\pi/a, 2\pi/a)$ (for the perpendicular components $p^{\perp} + p_2^{\perp} = p_1^{\perp} + p_3^{\perp}$). With the energy spectrum $\varepsilon(\vec{p})$ near the two opposite nodal points:

$$\varepsilon(\vec{p}) = \pm v_F(p^{\|} \mp p_F) + (p^{\perp})^2/2m, \quad (7)$$

it gives:

$$p^{\|} + p_2^{\|} - p_1^{\|} - p_3^{\|} - K^{\|} = v_F^{-1}[\varepsilon(\vec{p}) + \varepsilon(\vec{p}_2) + \varepsilon(\vec{p}_1) + \varepsilon(\vec{p}_3) - \Sigma] = 0. \quad (8)$$

In (8):

$$\Sigma \equiv \{(p^\perp)^2 + (p_2^\perp)^2 + (p_1^\perp)^2 + (p_3^\perp)^2\}/2m^* - 4\Delta. \quad (9)$$

Here $\Delta = v_F p_F [(K/4p_F) - 1] \equiv (v_F p_F)\kappa$.

It follows from (8) that the $\delta$-function $\delta(\varepsilon_1 + \varepsilon_3 - \varepsilon - \varepsilon_3)$ in (5) can be written in either of the two forms:

$$\delta(\varepsilon_1 + \varepsilon_3 - \varepsilon - \varepsilon_2) = (1/2)\delta(\varepsilon_1 + \varepsilon_3 - \Sigma/2), \text{ or } \delta(\varepsilon_1 + \varepsilon_3 - \varepsilon - \varepsilon_2) = (1/2)\delta(\varepsilon + \varepsilon_2 - \Sigma/2). \quad (10)$$

The right hand side (*R.H.S.*) in Eq.(5) linear in $\vec{u}$ with the help of Eq.(10) acquires the form:

$$R.H.S. = \iint |V(1;2)|^2 \frac{d^2\vec{p}_1 d^2\vec{p}_2}{T(4\pi)^4} \times \frac{(\vec{K}\cdot\vec{u})\delta(\varepsilon + \varepsilon_2 - \Sigma/2)}{ch(\varepsilon/2T)ch[(\varepsilon - \Sigma/2)/2T]ch(\varepsilon_1/2T)ch[(\varepsilon_1 - \Sigma/2)/2T]}. \quad (11)$$

(In (11): $|V(1;2)|^2 \equiv |V(\vec{p}_1\vec{p}_3;\vec{p}\vec{p}_2)|^2$). Since $d^2\vec{p}_1 d^2\vec{p}_2 = dp_1^\perp dp_2^\perp dp_1^\parallel dp_2^\parallel \equiv v_F^{-2} dp_1^\perp dp_2^\perp d\varepsilon_1 d\varepsilon_2$, one can directly integrate over $d\varepsilon_1$ in (11):

$$\int \frac{d\varepsilon_1}{ch[\varepsilon_1/2T]ch[(\varepsilon_1 - \Sigma/2)/2T]} = \frac{\Sigma}{sh(\Sigma/4T)}. \quad (12)$$

Combining (11) with the linear in electric field term in (4), $R.H.S. = ev_F (dn_{p,o}/d\varepsilon)$ and integrating both sides over $\vec{p}$ ($d^2\vec{p} \equiv v_F^{-1} dp^\perp d\varepsilon$) gives:

$$-eE \int_{\Delta p^\perp} dp^\perp = (Ku)T \iiint |V(1;2)|^2 \frac{dp^\perp dp_1^\perp dp_2^\perp}{4^2 \pi^4 v_F^3} \left(\frac{\Sigma/4T}{sh(\Sigma/4T)}\right)^2. \quad (13)$$

The integral in the left side is over the "arc width", $\delta = 2\Delta p^\perp = (\Delta\varphi)p_F$. The exact definition would depend on details of the arc's structure [15]. In what follows, $\delta$ is the model parameter.

At low temperatures the main contribution into the integral Eq. (13) comes from small $\Sigma$ of Eq. (9): $<(p^\perp/p_F)^2> \approx (K/4p_F - 1)$. Correspondingly, for $p^\perp$ to lie inside the arc's length $\delta$, the square root, $\kappa^{1/2} = \sqrt{K/4p_F - 1}$ has to be small enough. (For the Umklapp processes to be effective at low $T$ $\Delta$ must be positive: $\Delta = v_F p_F [(K/4p_F) - 1] > 0$).

Reduction of the quadratic in momenta term in $\Sigma$ (9) to the diagonal form simplifies the threefold integral in (13) to:

$$I \equiv \iiint \frac{dp^\perp dp_1^\perp dp_2^\perp}{(\pi)^4} \left(\frac{\Sigma/4T}{sh(\Sigma/4T)}\right)^2 = \frac{2\sqrt{2}}{\pi^3} \int_0^\infty \sqrt{v}\, dv \left[\frac{\{(v/4mT) - \Delta/T\}}{sh\{(v/4mT) - \Delta/T\}}\right]^2. \quad (14)$$

Then:

$$I = \frac{2\sqrt{2}}{\pi^3} \int_{-4m\Delta}^\infty \sqrt{(s + 4m\Delta)}\, ds \left[\frac{s/4mT}{sh(s/4mT)}\right]^2. \quad (15)$$

At low temperatures ($mT \ll \Delta$):

$$I \cong \frac{8p_F^3}{3\pi} \sqrt{\frac{\Delta}{\varepsilon_F}} \left(\frac{T}{\varepsilon_F}\right) \left(1 - \frac{\pi^2 T}{40\Delta} - ...\right) . \quad (16)$$

With matrix elements $|V(1;2)|^2 \equiv |V(\vec{p}_1\vec{p}_3; \vec{p}\vec{p}_2)|^2$ in the dimensionless form: $|V(1;2)|^2 = |\tilde{V}(1;2)|^2 \, v_0^{-2}(\varepsilon_F)$ (here $v_0(\varepsilon_F) = (m/2\pi\hbar^2)$ is the "bare" 2D density of states for one spin direction) one has $eE\delta = up_F (|\tilde{V}(1;2)|^2 / v_F^3) T(\pi/m)^2 I$; that gives for the drift velocity:

$$u = \frac{-eE\delta v_F^3}{|\tilde{V}(1;2)|^2 \, T p_F (\pi/m)^2 I} . \quad (17)$$

Substitution of $n_1(\vec{p})$ in its form Eq.(6) into the expression for current, $j_{2D} = 4e\int v_F \frac{d^2\vec{p}}{(2\pi)^2} n_1(\vec{p})$ after trivial transformations and using that $d^2\vec{p} \equiv v_F^{-1} dp^\perp d\varepsilon$ gives the in-layer resistance:

$$\rho_{2D}(T) = \left(\frac{\pi\hbar}{e^2}\right) \frac{4\pi^2 |Z\tilde{V}(1;2)|^2}{3(\Delta\varphi)^2} \sqrt{\frac{\Delta}{\varepsilon_F^*}} \left(\frac{T}{\varepsilon_F^*}\right)^2 . \quad (18)$$

(Electrons on the two opposite arcs contribute equally into the current, hence, the additional factor 2 in $j_{2D}$ above beside spin).

The $T^2$ dependent resistivity, Eq. (18) is the central result of the paper. Notice that renormalization of the interaction, $\tilde{V}(1;2) \to Z\tilde{V}(1;2)$ and of the mass, $m* = Z^{-1}m$ is already taken into account in Eq. (18). The combination $\varepsilon_F^* = p_F^2 / 2m^*$ is the renormalized band Fermi energy; here and below $\varepsilon_F^*$ is the convenient shorthand notation for the actual expansion parameter.

Let us do some estimates. First, taking , $m^* \approx 4m$ and the lattice parameter $a \simeq 3.86 \cdot 10^{-8} cm$ for YBCO one would obtain $\varepsilon_F^* \approx 3000K$ and $T^{**}/\varepsilon_F^* \approx 1/15$. (For comparison, the Fermi energy of Tl2201 estimated from data on QO [19] is $\varepsilon_F \approx 5000K$).

In Eq. (18) $\Delta/\varepsilon_F^* = 2|K/p_F - 1| \equiv 2\kappa$. Substitution of the same data for YBCO (p=0.09) [4] gives for the arc the reasonable estimate: $\Delta\varphi \approx 0.37\kappa^{1/4} \approx 21°\kappa^{1/4}$ (assuming $|Z\tilde{V}(1;2)|\sim 1$). There, only ARPES can determine positions of the nodal points in the BZ, however, even if $\kappa$ were small, $\kappa^{1/4} \sim 1$.

In the theoretical formulas they deviations from the quadratic T-dependence Eq. (18) occur at $T \approx (40/\pi^2)\Delta \sim 8\kappa\varepsilon_F^*$ (see Eq. (16)). Such value $T^{**}$ would seem too high. In the estimates above it was implicitly assumed that in actual fact, $T^{**}$ is due to switching on other mechanisms, say, scattering on the phonons or the pseudogap diminishing at $T \approx 200K$ [4,13].

In regard to the $1/p$-dependence on $p$, the dopants' concentration for the FL resistivity in Hg2201 [4], notice the factor $(\Delta\varphi)^2$ in the denominator of Eq. (18): if the arcs have had two sides, $(p_F\Delta\varphi)^2$ would be proportional to "area" of the "enclosed pocket".

Turn to Hall Effect. In non-zero magnetic fields, $H$ the Lorentz force causes the flow of excitations perpendicular to the electric field on arcs on the second diagonal. Calculations of the transverse drift velocity are identical to the ones for conductivity. The Hall coefficient $R_H$ is:

$$R_H = \frac{\pi^2}{ec(\delta p_F)}. \quad (19)$$

In literature FS of UD cuprates is routinely called as the hole-like FS centered at $(\pi,\pi)$. In semiconductors the concept of holes emerges at the chemical potential being close to the top of an electronic band. This point of view is not applicable to UD cuprates because the underlying FS is not small. Instead, since the system with the electronically half-filled BZ is in the insulating Mott state, I argue that, counting off from the insulating limit, the carriers on the Fermi arcs are holes, $e > 0$ and $R_H > 0$. (In Eq. (19) $\delta p_F / \pi^2 = \Delta\varphi(p_F^2/\pi^2)$ stands for the effective number of carriers in the Fermi-arcs).

Derivation of Eq. (19) assumes the classical regime $\omega_c \tau << 1$ ($\omega_c = (eH/m^*c)$). Peculiarity of the situation is that by the order of magnitude $\tau$ typically is: $\tau \propto \varepsilon_F / T^2$, i.e., is very large in clean samples. Therefore, at $T = 100K$, for example, the magnetic field needs to be even smaller than $3-5 Tesla$. Data for $R_H$ in Hg1201 and YBCO in same temperature range are represented only for fields exceeding $28 Tesla$ (see [20] and references wherein).

In the context of the model one can suggest another interpretation to the provision $\omega_c \tau << 1$. The transverse momentum component obeys the equation: $\partial p^\perp / \partial t = (ep_F / cm^*)H$. Then the deviation $\Delta p^\perp \simeq \omega_c H \tau$ in the transverse direction for the time-period $\tau$ should not exceed the arc's length $\delta$.

Finally, a few comments about what is known concerning the paramagnetic phase in UD cuprates below the pseudogap temperature $T^*: T_c < T < T^*$. In the introductory part it was argued that experiments [3] disprove the case of the Fermi surface reconstructed via a spin- or charge- wave transition in UD *ortho II* YBCO. For Hg2201 and YBCO there are experimental evidences in favor of the transition into the so-called "loop currents phase "(see [21] and the experimental reference wherein). By its symmetry [22], this order parameter would only break the tetragonal symmetry in positions of the pairs of the nodal points in the BZ. Therefore one would expect no other changes in the above results except an anisotropy in the $T^2$-resistivity contribution.

(An uncertainty regarding structure of electronic spectrum in the pseudogap phase, however, was brought about recently by observation of the *two* phase transition in (nearly) *optimal doped* single-layer Bi2201 at $T = T_c$ and $T = T^*$ [23]).

To conclude, for temperatures above the superconducting dome the analysis of ARPES data [10-13] made it possible for us to develop the model in which carriers in the Fermi arcs in the normal state occupy narrow valleys in the momentum space restricted on each side by regions with the large energy gap. Inside each valley the excitations' spectrum is almost the ordinary energy spectrum of free holes. Strong short-range electron-electron interactions are assumed to prevail in the relaxation processes in the sub-system of the Fermi arcs for clean UD cuprates and at low temperatures. Scattering of carriers between four arcs at the nodal points in the BZ was treated in frameworks of the appropriate kinetic equation. After that, the FL renormalization owing to the electron-electron correlations was taken into account for all parameters. The explicit treatment of the Umklapp processes resulted in the finite conductivity value. The resistivity of the normal phase calculated at low enough temperatures is proportional to $T^2$, in the excellent agreement with the experimental data [4]. It is shown that Hall Effect in clean UD cuprates is extremely sensitive to the value of magnetic fields at these temperatures. Presumably, even the sign of the Hall coefficient may depend on the magnetic field strength. Experimental data for the Hall coefficient for low magnetic fields are missing.

## Acknowledgements

The author is very grateful to N. Barisic for bringing the interesting article [4] to his attention. I am also thankful to O. Vafek for many stimulating discussions of the problem and to I. Kupcic and S. Barisic for sharing their unpublished theoretical results on the magneto-transport in HTSC.

The work was supported by the NHMFL through NSF Grant No. DMR-0654118 and the State of Florida.